\documentclass[aps,pra,showpacs,superscriptaddress, twocolumn]{revtex4-1}%
\usepackage{amsmath}
\usepackage{amssymb}
\usepackage{amsfonts}
\usepackage{float}
\usepackage{graphicx}
\usepackage[colorlinks,linkcolor=blue,anchorcolor=blue,citecolor=blue,urlcolor=blue]{hyperref}
\usepackage{mathrsfs}
\usepackage{epsfig}
\usepackage{subfigure}
\usepackage{CJK}

\begin{document}
\begin{CJK}{UTF8}{gbsn}
\title{Spin readout of trapped electron qubits}
\author{Pai Peng (彭湃)}
\affiliation{{Department of Physics, University of California, Berkeley, California 94720, USA}}
\affiliation{{School of Physics, Peking University, Beijing 100871, People's Republic of China}}
\author{Clemens Matthiesen}
\author{Hartmut H\"{a}ffner}
\email{hhaeffner@berkeley.edu}
\affiliation{{Department of Physics, University of California, Berkeley, California 94720, USA}}

\date{\today}

\begin{abstract}
We propose a scheme to read out the spin of a single electron quantum bit in a surface Paul trap using oscillating magnetic field gradients.
The readout sequence is composed of cooling, driving, amplification and detection of the electron's motion. We study the scheme in the presence of noise and trap anharmonicities at liquid helium temperatures. An analysis of the the four procedures shows short measurement times ($25~\mu$s) and high fidelities ($99.7\%$) are achievable with realistic experimental parameters.
Our scheme performs the function of fluorescence detection in ion trapping schemes, highlighting the potential to built all-electric quantum computers based on trapped electron spin qubits.
\end{abstract}

\maketitle
\end{CJK}

\section{Introduction}
Trapped charged particles \cite{RMP1990Paul, RMP1990Dehmelt} are promising candidates for the implementation of quantum information processing (QIP) schemes \cite{DiVincenzo, QCs, RMP-QC-Shor}.
Ions, in particular, have been studied extensively during the past two decades \cite{PRL1995CiracZoller, PRL1995Wineland, RMP1999Wineland,RMP2003Leibfried,Haeffner2008, Science2016Chuang}.
Both motional and electronic quantum states of trapped ions can be controlled extremely well with laser light . In addition, they possess well-isolated states whose long lifetimes and coherence times are suitable for storage of quantum information \cite{IEEE1991Wineland, PRL2005Wineland, JPB2003Blatt}.
Entangling two-qubit gates commonly rely on the motional degree of freedom as quantum bus between the ions \cite{PRL1995CiracZoller, PRL1999SorensenMolmer, Nature2003Blatt} and with typical gate times of 10-100 $\mu$s constitute a bottleneck in terms of computation speed. Further, the use of lasers for state initialization, cooling, readout, and qubit control leads to a large experimental overhead which poses a challenge to scaling up the current technology to many qubits \cite{QIC2007Steane}.
Trapped electrons could provide an attractive alternative to atomic ions. Their spin states can encode the qubit while spin-motion coupling in a harmonic trapping potential would mediate coupling to neighboring electrons. Importantly, electrons are four orders of magnitude lighter than the atomic ions used in QIP experiments, offering the potential to speed up two-qubit gates \cite{PRL2003Tombesi, PRA2010Gabrielse}.

Trapped electrons can benefit from the technology developed around trapped ions. Notably, scalable QIP architectures like the quantum CCD-architecture based on ion shuttling \cite{Nature2002Wineland} can be adopted without conceptual changes. Guiding electrons along a microfabricated quadrupole waveguide has already been demonstrated \cite{PRL2011Hommelhoff,PRL2015Hommelhoff} and shuttling electrons in segmented traps appears straightforward.  
Further, single-qubit gates for electron spins, as for hyperfine states of ions, can be realized using transverse radiofrequency fields. Two-qubit microwave gates for ions \cite{Nature2011Wineland,PRL2001Mintert,Nature2011Wunderlich} can also be adapted to trapped electrons with the added benefit that these gates will speed up considerably due to the small mass of electrons.
Proposals towards QIP with individual trapped electron spins have been developed for  Penning traps \cite{PRL2003Tombesi,JPB2009Marzoli,PRA2010Gabrielse} and more recently for Paul traps \cite{NJP2013Haeffner,arXivWineland}.

However, adapting the ion-trap blueprint to electrons is complicated by the lack of optical transitions and fast spontaneous emission channels, which appear to pose serious challenges for electron spin initialisation and readout.
In previous Penning trap experiments aimed at measurements of the $g$-factor of the electron, spin readout of trapped electrons was achieved using the so-called continuous Stern-Gerlach effect \cite{RMP1986, PNAS1986a, PRL2000Werth}, but spin detection times on the order of seconds render such schemes unattractive for use in QIP.
Another option would be to couple the electron to solid-state quantum systems with the promise of sub-microsecond readout times, but a satisfactory interface between charged particles and solid-state systems has not been realized as of yet \cite{NJP2013Haeffner, ARCMP2013Haeffner,arXivWineland}.

Here, we propose a scheme to read out the spin state of a single electron trapped in a linear surface trap under a static magnetic field without the assistance of additional quantum systems. For ease of experimental implementation, we design and study this scheme in view of its compatibility with the rather modest cryogenic requirements of a 4-K environment as well as planar Paul traps. We note that readout-conditioned single-qubit rotations can be used to initialize the spin-qubit in a well-defined state. Hence, with state readout, the trapped electron platform discussed here satisfies the DiVincenzo criteria \cite{DiVincenzo} for QIP and enables a trapped electron architecture very similar to current trapped ion approaches \cite{Wineland1998,Nature2002Wineland}. Thus, we can hope to combine the advantages of trapped ions, namely that of a flexible architecture and long memory times, with those of high speed gate operations due the small mass of the electrons and robust electronic control.

At the core of our proposal for state readout is a dynamic version of the Stern-Gerlach effect where the electronic motion is driven with an oscillating linear magnetic field gradient resonant with the secular frequency of the trapped electron.
Since the force is proportional to the projection of the electron spin onto the direction of a static magnetic field, spin-up and spin-down electrons experience opposite forces thereby creating motion with opposite phases. The electron's motion induces an image current in pick-up electrodes on the surface trap, the phase of which is then amplified and measured electronically. The experimental challenge consists of exciting a large enough state-dependent coherent motion to overcome noise originating from the original random thermal motion of the electron and the Johnson noise of the detection circuit.
We expect Johnson noise in the electronics to be the larger contribution as the electron motion can be cooled with adiabatic or parametric coupling schemes below the Johnson noise limit of an attached cooling circuit as already demonstrated in Penning trap experiments \cite{vanDyck1978electron}.

\begin{figure}\centering
\includegraphics[width=0.45\textwidth]{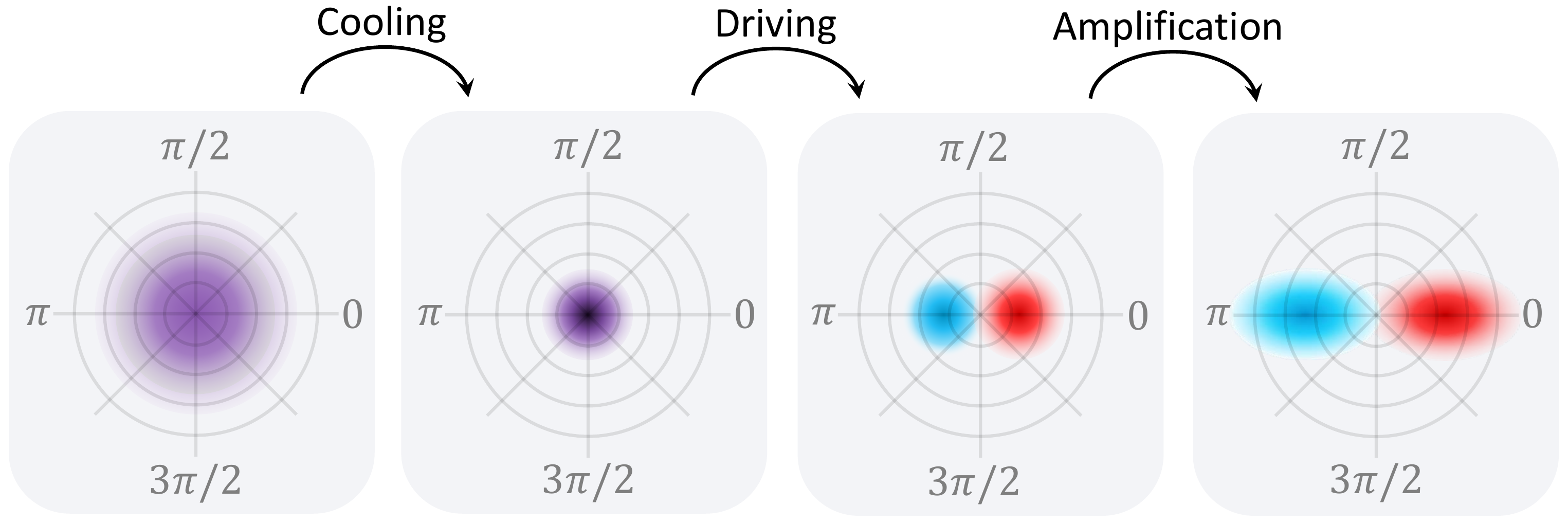}
\caption{(color online) Illustration of the spin readout sequence for a single electron at 4-K temperature. The initial amplitude and phase of the electron's axial motion is described by a probability distribution function (PDF) of the amplitude $A_\phi$, where the phase $\phi$ is relative to the driving field. Cooling the electron motion leads to a narrower probability distribution. Then, by driving the electron with a spin-dependent force from the magnetic field gradient, the two PDFs associated with the respective spin eigenstates separate along the $\phi=0$ axis. The blue (red) area corresponds to the PDF for the electron in the spin up (down) state. Amplifying both the coherently driven motion and the thermal motion allows detection of a signal by a resonant detection circuit at 4-K temperature.}
\label{fig:flowchart}
\end{figure}

Several experimental steps are necessary to achieve high-fidelity state readout: First, an individual mode of the electron motion is cooled from the environment temperature of 4~K to $\sim$ 0.4~K. This is achieved either by parametrically coupling the mode to another high-frequency mode which in turn is resonant with an LC resonator \cite{vanDyck1978electron,RMP1986} at 4~K, or by first thermalizing the mode of interest with the resonator at a higher trap frequency, and adiabatically lowering its frequency by one order of magnitude. In the next step, alternating currents produce a spin-dependent magnetic force, creating a coherent state with an amplitude exceeding the amplitude of the initial thermal motion. As trap anharmonicities might wash out the phase relation between the drive and the electron motion, it is critical to keep this step short. However, the coherent state amplitude needs to exceed the detection limit posed by the Johnson noise of the detection electronics. Hence, a second process using stronger electric forces instead of magnetic forces is used to amplify the motion more rapidly while preserving phase information. This can be achieved via parametric amplification of the motion by modulating the curvature of the trapping field at twice the mode frequency \cite{PRD1982Caves, PRL1991Gabrielse,APB1995Gabrielse} up to the point where the coherent state amplitude exceeds the thermal noise of the detection circuit. Finally, the mode frequency is tuned into resonance with the detection circuit, thereby allowing readout of the spin state via the phase of the detected image current. Figure~\ref{fig:flowchart} illustrates the axial motion probability distribution of the electron's amplitude and phase at different stages of the above protocol.
Analysis and optimization of the above procedure yields an estimate of the readout fidelity of $99.7\%$ with $25~\mu$s total measurement time. In comparison, the spin coherence time of the electron in a surface Paul trap is expected to be in the seconds regime \cite{APB2016Ruster}. More ambitious cooling techniques or simply lowering the base temperature of the cryostat below 1~K allows for further improvements in detection fidelity and speed.

We next discuss the details of the considered experimental system in section \ref{system}. Section \ref{procedure} describes the readout procedure and provides numerical results. Our work is summarized in section \ref{conclusion}.

\section{Characterization of the system}\label{system}

\begin{figure}\centering
\includegraphics[width=0.45\textwidth]{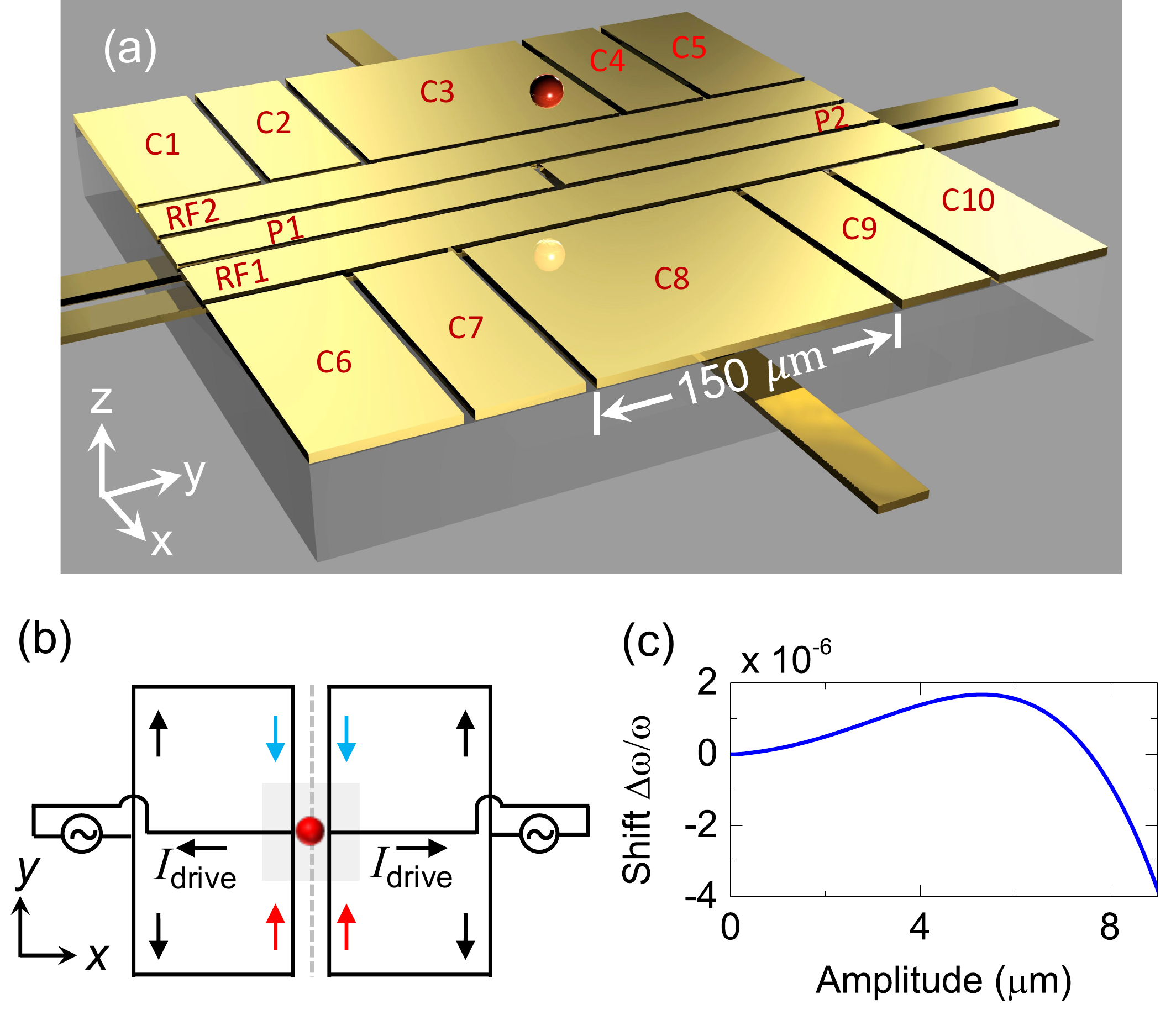}
\caption{(color online) (a) Central portion of the surface-electrode trap design for electron spin readout. The electron is trapped 33~$\mu$m above the electrode surface. Electrodes C1-C10 provide an axial confinement potential, while the radiofrequency electrodes RF1 and RF2 create the radial confinement. The central electrode (P1, P2)  is split to detect the image current $I_\mathrm{image}$ induced by the axial motion about the trap center. Current-carrying wires under the electrode surface produce an oscillating magnetic field gradient. (b) Top view of the circuit which produces the oscillating magnetic field gradient. Arrows indicates the direction of driving current $I_\mathrm{drive}$ at one instance in time. The wire sections marked by blue and red arrows generate the field of interest while the black ones illustrate current flow in the other parts of the circuit. The vertical grey dashed line represents the axial direction and the red sphere marks the trapping position. The central portion of the trap, which is shown in (a), is marked by the shaded area. (c) Shift of the axial frequency as a function of the motion's amplitude due to anharmonicities in the trap potential.}
\label{schematic}
\end{figure}

Our scheme assumes a surface Paul trap, which allows for scalable quantum computing architectures \cite{Nature2002Wineland}. The chip design, shown in Fig. \ref{schematic} (a), uses a typical ``five-wire'' configuration for trapping in the top layer.
Two radiofrequency (RF) electrodes produce the transverse confinement, trapping the electron at $h=33~\mu$m above the surface.
The distance is chosen as a compromise of two factors: (i) in order to generate a strong magnetic gradient at the electron position and to induce a large image current $I_\mathrm{image}$ a short distance is required; (ii) to avoid anomalous heating \cite{PRA2016Haeffner, PRB2014Haeffner, PRL2012Wineland, PRA2015Chiaverini} and to accommodate for the extent of the electron's thermal motion ($\approx \mu$m), the electron cannot be trapped too close to the surface.

A static magnetic field in the $x$-direction provides the quantisation axis. The precise splitting of the qubit states is not crucial for this proposal, but for concreteness, a frequency of 10-100 MHz seems appropriate, being both experimentally undemanding and sufficiently different from the secular trap frequency.
Ten direct current (DC) electrodes on both sides of the RF electrodes offer enough degrees of freedom to form a harmonic axial potential. A second layer of the trap chip contains the current-carrying wires which produce the magnetic field gradient to generate the spin-dependent force. However, the alternating current $I_\mathrm{drive}$ in the wires induces not only a magnetic field, but also an electric field. For the recent implementation of magnetic gradient microwave gates for ions employing a straight three-wire configuration \cite{Nature2011Wineland}, where we estimate the spin-independent electric force to be three orders of magnitude stronger than the magnetic force. As this strong electric force is on resonance with the axial motion, it would make it difficult to extract information on the spin direction of a trapped electron.
 
To suppress the axial electric field while maintaining a large magnetic gradient, we propose a symmetric design, see Fig. \ref{schematic}(b), which contains two circuits mirrored along the axial direction where ideally both the magnetic field and the axial component of the  electric field vanish at the center. Only the current marked by blue and red arrows contributes to the axial magnetic field gradient: if the electron moves in the $+y$ direction, it experiences a stronger magnetic field from the blue current, corresponding to a net magnetic field along the $-x$ direction, and vice versa. Assuming the wires are made of gold, have cross-sections of $1~\mu\mathrm{m}\times 10~\mu\mathrm{m}$, the parallel sections of the two circuits are separated by 20~$\mu$m, and a 1~A current is running in each circuit, we find the electron should feel a magnetic field gradient of 91~T/m at the center of the trap (Appendix \ref{field}). Shielding due to the 1~$\mu$m-thick electrodes located in the top layer 1~$\mu$m above the wire has been taken into account and reduces the magnetic field by about a third. Compared to previous work using three wires, where careful balancing of amplitudes and phases between the wires allowed canceling the magnetic field at the ion location while achieving a gradient of 35~T/m \cite{Nature2011Wineland}, the symmetric wire design proposed here should be experimentally robust and undemanding.

Deviations from the ideal design break the symmetry and lead to a non-zero electric force, but we expect this contribution to be at most on the same order as the magnetic force.
The effect of a net electric field can be reduced with a spin-echo sequence. When periodically changing the phase of the magnetic field drive while flipping the spin, the effect of the electric field cancels, while the force due to magnetic gradient continues to be in phase and the state is further displaced (Appendix \ref{imperfections}).
Although, in principle, dynamic decoupling can also cancel the electric field for the three-wire configuration, it would require an unrealistically high number of spin flips to ensure that the amplitude of the coherent state does not exceed the limits imposed by the residual trap anharmonicities discussednext.

Creating a harmonic axial potential is crucial for electron spin detection, as anharmonicities in combination with thermal motion lead to frequency broadening thereby washing out the phase information required for detection of the electron motion \cite{EPJD2008Bushev} and the spin direction. To assess the achievable degree of harmonicity, we calculate the axial potential using a simplified trap structure where the gaps between electrodes are infinitesimally small and optimize the voltages on DC electrodes numerically. We believe including gaps will not limit the achievable degree of harmonicity.
We assume voltages on each electrode are provided by conventional $\pm$10~V, 16-bit digital-to-analog converters, limiting the voltage resolution in this optimisation step. Expanding the optimized trap potential into the Taylor series $V(y)=V(0)(c_2 y^2+c_4 y^4+c_6 y^6)$, and $c_2=1 \left(\mu\mathrm{m}\right)^{-2}$ we find coefficients $c_4=10^{-7}\left(\mu\mathrm{m}\right)^{-4}$ and $c_6=-2\times 10^{-9}\left(\mu\mathrm{m}\right)^{-6}$, while odd and higher even order terms are negligible. The relative frequency shift as a function of the axial motion (Fig. \ref{schematic} (c)) is determined by $\Delta \omega/\omega\approx (3A^2c_4/4+15A^4c_6/16)/c_2$, where $A$ is the amplitude of the motion.
We find that anharmonicities can be suppressed such that the relative frequency shift is less than $10^{-6}$ for the few micron amplitudes we are interested in.

\section{Readout procedure}\label{procedure}

In this section, we present the detailed protocol and discussion for the cooling-driving-amplification-detection procedure.

\subsection{Cooling}\label{cooling}
The timescale of interest for the readout scheme is less than 100~$\mu$s, which is short as compared to the expected timescales for anomalous heating \cite{RMP2015Brownnutt}. Further, until the start of the detection phase of the scheme the detection circuit is also assumed to be detuned from the axial motional frequency, avoiding heating of the motion. 
The detuning can be realized by either tuning the secular trap frequency or using a tunable capacitance in the detection circuit.

The first part of our protocol aims to cool the axial thermal motion such that the magnetic driving force dominates the motion after a reasonable time. One cooling method is parametric swapping of the population between the axial mode of frequency $\omega$ and a transverse mode of higher frequency $\omega_\mathrm{t}$ \cite{vanDyck1978electron,PRAparametriccoupling}. We assume that initially both modes are at the environmental temperature $T_\mathrm e=4$~K with their populations determined by the Boltzmann distribution. Then, after population exchange the temperature of the axial mode is cooled to $T_0=T_\mathrm e \omega/\omega_\mathrm{t}$. For an ion in a surface trap exchange times on the order of $100~\mu$s have been achieved \cite{PRAparametriccoupling}. Due to the linear mass dependence, we expect parametric swapping to take place in less than a microsecond for electrons. An alternative cooling method consists of adiabatically lowering the axial frequency from $\omega_0$ to $\omega$, such that $T_0=T_\mathrm e\omega/\omega_0$ \cite{arXivDrewsen}. Adiabaticity is satisfied for frequency ramps on timescales slower than $0.5/(2\pi\omega)$. For example, with $\omega=2\pi\times300$~MHz a 100-ns ramp is well adiabatic. A transverse secular frequency of $\omega_\mathrm{t}=3$~GHz in the parametric scheme or a $\omega_0=3$~GHz initial axial secular frequency in the adiabatic scheme should allow cooling the axial motion to $T_0=0.4$~K, which we assume to be the temperature of the electron motion for the  following steps.

\subsection{Driving}

The goal of driving is to separate the axial motion of spin-up and spin-down electrons, which obey the same axial distribution function at $T_0$ after cooling (see Fig. \ref{fig:flowchart}). The separation can be realized by a spin-dependent force that maps the spin state to the axial motion. In the presence of an oscillating magnetic field gradient, the axial Hamiltonian can be written
\begin{equation}
\hat H=\frac{\hat{p_y}^2}{2m}+\frac{1}{2}m\omega^2 \hat{y}^2 + \sum_{n=3}^{\infty} V_n \hat{y}^n -\mu_\mathrm B B_x'(y)\hat{\sigma_x}\hat y e^{i\omega t},
\end{equation}
where $\hat y$ ($\hat{p_y}$) is the position (momentum) operator, $m$ is the mass of electron, $\omega$ is the axial trap frequency and $V_n$ is the $n$th-order expansion coefficient of the potential, accounting for anharmonicity. The last term arises from the resonant driving, $B_x'(y)$ the derivative of magnetic field along $x$ with respect to $y$, $\mu_\mathrm B$ the Bohr magneton, and $\hat{\sigma_x}$ the Pauli operator along the $x$ direction. As the qubit is projected into an eigenstate of $\hat{\sigma_x}$ during the measurement, the operator can be replaced by a scalar $\sigma_x=\pm1$.
Defining the annihilation operator as in the case of a harmonic oscillator we obtain the Heisenberg equation of motion (in the frame rotating with frequency $\omega$)
\begin{equation}\label{EOM}
i\hbar \frac{ \mathrm d\hat a}{\mathrm dt}=-\sum_{n=3}^{\infty} n V_n y_0^n (\hat{a}+\hat{a}^\dag)^{n-1}+B_x'(y)\mu_\mathrm B{\sigma_x}y_0,
\end{equation}
where $y_0=\sqrt{\hbar/(2m\omega)}$ is the ground state extension of the harmonic oscillator.
The trap frequency is set to be $\omega=2\pi\times300$~MHz.
At 0.4~K, $\langle\hat{a}^\dag \hat {a}\rangle\approx 30\gg1$, so we can ignore all commutators and replace the annihilation operator $\hat a$ by its expectation value $\langle \hat a \rangle$, whose real (imaginary) part represents the amplitude $A_0$ ($A_{\pi/2}$) in phase (in quadrature) with the drive.
As spin-up and spin-down electrons feel opposite forces, the spin information is encoded in the amplitude of the in-phase motion $A_0$, whose sign $\mathrm{sgn}(A_0)$ determines the measurement result (spin up or spin down), and whose modulus $|A_0|$ determines the size of the signal. The in-quadrature motion is not directly related to the spin state, so we focus on the in-phase motion in the following discussion.
To gain analytical insight, we first assume a perfectly harmonic trap potential, so the initial thermal motion and the driven motion are independent.
The thermal distribution of the in-phase amplitude (or, equivalently, the probability distribution of the axial position of the electron at $t=0$) is described by a Gaussian distribution with zero mean and standard deviation $\sigma_\mathrm{std}=\sqrt{k_\mathrm BT_0/m\omega^2}\approx 1.3~\mu$m.
The in-phase amplitude of a resonantly driven oscillator grows linearly with time at a rate $B_x'(y)\mu_\mathrm B{\sigma_x}/2m\omega$, so a $\sigma_x=1~(-1)$ electron acquires a positive (negative) in-phase amplitude $A_0$ after driving.
To achieve high readout fidelities the axial probability distributions along $\phi=0$ for the two spin populations should be well separated after driving (see Fig. \ref{fig:flowchart}). Here we aim for the average amplitude of the axial motion for each spin state after driving to be larger than $3\sigma_\mathrm{std}$.

\begin{figure}\centering
\includegraphics[width=0.5\textwidth]{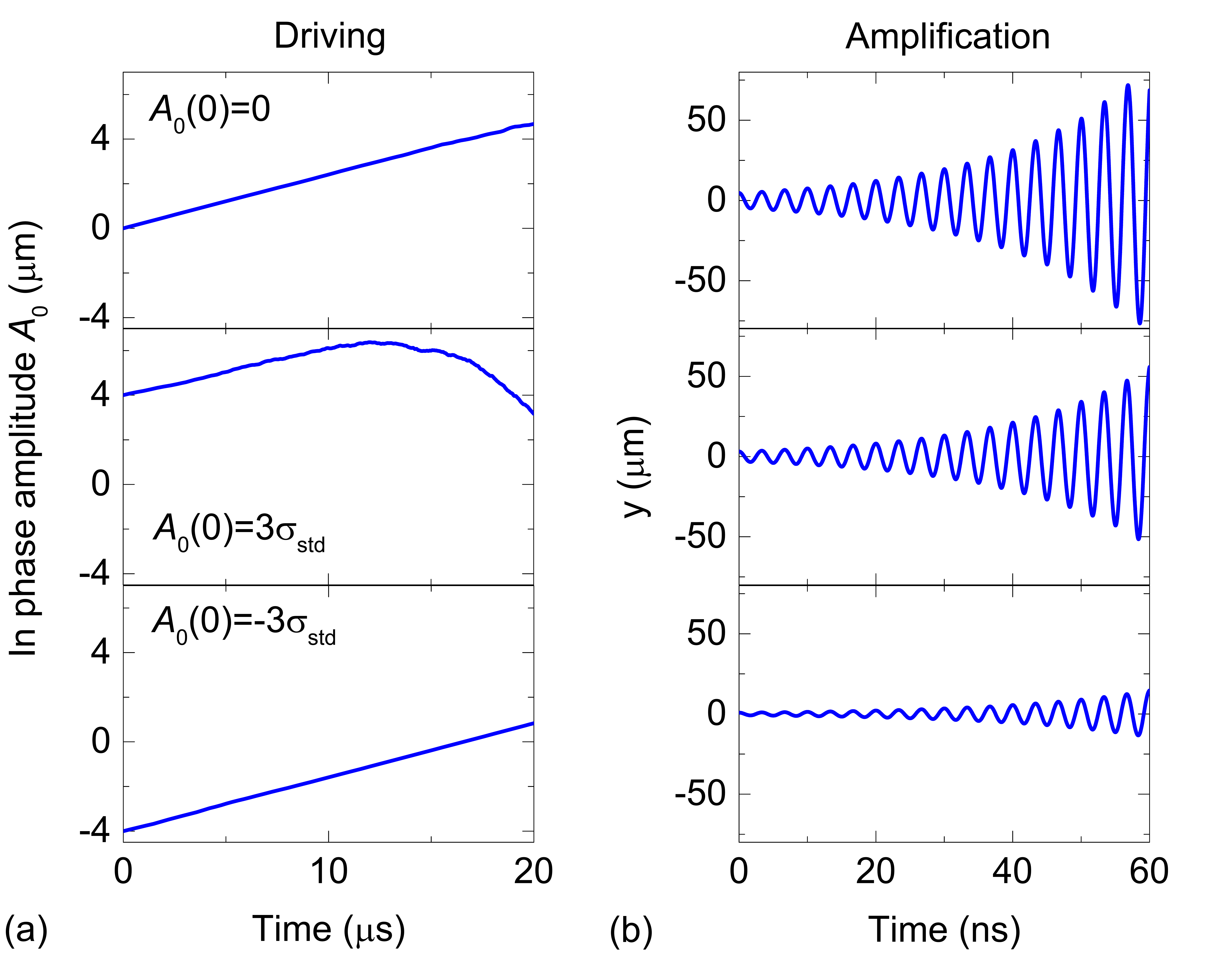}
\caption{(color online) (a) In-phase amplitude of the axial motion during the driving process of a spin-up electron with three initial axial locations: 0 and $\pm 3\sigma_\mathrm{std}$, while $A_{\pi/2}=0$ for all three panels. (b) Axial motion during the amplification process corresponding to the three cases in (a).}
\label{fig:time_evo}
\end{figure}

To support our analytical argument above, we simulate the electron motion based on Eq. (\ref{EOM}).
The Taylor expansion coefficients $V_n$ are extracted from a fit to the optimized potential obtained in Sec. \ref{system} in the axial range (-10~$\mu$m, 10~$\mu$m). Anharmonic terms are retained to $8^{\mathrm{th}}$ order in $y$, and terms due to the oscillating magnetic field gradient to $4^{\mathrm{th}}$ order.
As the axial potential is symmetric, electrons with spin up and spin down are connected by a transformation $y\to-y$, so we consider the spin-up case only. Figure \ref{fig:time_evo}(a) shows the in-phase amplitude for electrons with different initial axial positions: $A_0(t=0)=0$, $3\sigma_\mathrm{std}$ and $-3\sigma_\mathrm{std}$. $A_{\pi/2}(0)=0$ for all the three cases. The expected result for spin up is $A_0>0$ after driving, and the three cases show no exception. However, a drop of $A_0$ is observed for $A_0=3\sigma_\mathrm{std}$, because its amplitude exceeds $6~\mu$m during the driving process. For large amplitudes the anharmonicity causes a shift of the axial frequency and the electronic motion will be out of phase with the drive after some time, thereby decreasing the in-phase amplitude. To avoid strong dephasing for large initial amplitudes, we keep the driving time to be as short as possible. 
The driving force is then just sufficient to provide an average $A_0$ slightly larger than $\sigma_\mathrm{std}$ as shown in Fig. \ref{fig:time_evo}(a).
 In addition, the figure also highlights the necessity of the cooling process: the magnetic driving itself would not be strong enough to separate spin-up and spin-down electrons with a $T_\mathrm{e}=4$~K thermal distribution.

\subsection{Amplification}

The spin-dependent motion dominates the $T_0=0.4$ K thermal motion after driving, but it is still weak compared with Johnson noise at $T_\mathrm e=4$ K. Therefore the next step in the protocol is parametric amplification, with the intention to magnify the electron's axial motion, both coherent and incoherent parts. The amplification is realized by amplitude-modulation of the trap potential such that $V=V(0)(1+\epsilon\sin2\omega t)$, where $\epsilon\ll1$ is the modulation amplitude \cite{PRD1982Caves,PRL1991Gabrielse,PRA1999Gabrielse}. Parametric amplification leads to an exponential increase (decrease) of the amplitude in phase $A_0$ (in quadrature $A_{\pi/2}$) at a rate $\epsilon\omega/4$. Therefore the spin information (sign of $A_0$) is conserved during the amplification process. As the amplification is achieved by electric fields, much stronger drives as compared to the magnetic gradients are feasible. Since the phase difference is the approximate product of frequency shift and time, even though the anharmonicity-induced frequency shift is large when the motion is amplified, dephasing is not evident. For $\epsilon=0.1$, taking into account nonlinearities up to 8$^\mathrm{th}$ order from a fit to the potential in the range $(-100~\mu\mathrm{m},~100~\mu\mathrm{m})$, we simulate the motion of an electron during the 60-ns amplification process for the three cases shown in Fig. \ref{fig:time_evo}(a) and plot them in Fig. \ref{fig:time_evo}(b). The amplitude can be amplified to 50~$\mu$m without loss of phase information.

\subsection{Detection}

The final step consists of bringing the detection circuit and the electron motion into resonance and detecting the image current $I_\mathrm{image}$.
Detection is achieved by connecting the two ends of the split 30~$\mu$m-wide central electrode to an external circuit and measuring the phase of $I_\mathrm{image}$ using phase-sensitive detection.
The local oscillator in this scheme is set to be in phase with the motion of spin-up electrons and in the opposite phase with spin-down electrons.
We can neglect the effects of anharmonicity in the signal-to-noise analysis as the detection bandwidth is orders of magnitude larger than motional broadening due to anharmonicity. Johnson noise, modeled as white noise, is the dominant noise noise during detection.
Following Refs. \cite{RMP1986, PNAS1986b}, the signal-to-noise ratio is obtained as (see Appendix \ref{stn} for more detail)
\begin{equation}\label{eq:stneq}
\frac{S}{\sqrt{\langle N^2\rangle}}=\sqrt{\frac{m\omega^2 A_\mathrm{0,amp}^2}{\gamma t_\mathrm{det}k_\mathrm B T_\mathrm e}},
\end{equation}
where we assume $\exp(-\gamma t_\mathrm{det})\ll 1$, so that the electron motion is completely damped, i.e. the spin information is fully transferred from the axial motion to the detector. $S$ ($N$) is the signal (noise) voltage,  $A_\mathrm{0,amp}$ is the in-phase amplitude after amplification, $\gamma$ is the damping rate due to the interaction with the detection circuit, $t_\mathrm{det}$ is the detection time and $k_\mathrm B$ is the Boltzmann constant. The harmonic motion of the electron forms an effective $LC$ circuit resonant at $\omega$. The effective inductance $L_\mathrm{eff}$ depends solely on the geometry of the electrodes with respect to the electron location, and the effective capacitance is $C_\mathrm{eff}=(\omega^2 L_\mathrm{eff})^{-1}$. Here, $L_\mathrm{eff}=0.15$~H and $C_\mathrm{eff}=1.9$ aF with $\omega=2\pi\times300$~MHz. The damping rate is determined by $\gamma=R/L_\mathrm{eff}$, where $R$ is the real part of the impedance of detection circuit. The right-hand side of Eq. \ref{eq:stneq} is proportional to the square root of the ratio of the electron's motional energy to the thermal energy. The signal-to-noise ratio scales as $t_\mathrm{det}^{-1}$ because the amplification is off during the detection, so the signal decays exponentially while noise is present throughout the process. To obtain a large signal-to-noise ratio as well as to guarantee that the electron motion is fully damped, we choose the detection time to be $t_\mathrm{det}=4\gamma^{-1}=4~\mu$s, for a detection circuit with on-resonance resistance 160~k$\Omega$. Equation \ref{eq:stneq} explicitly demonstrates the need for amplification: the coherent motional energy after the magnetic gradient drive is slightly larger than 0.4~K, and hence this signal is too weak to be detected in the presence of 4-K Johnson noise.

\begin{figure}\centering
\includegraphics[width=0.45\textwidth]{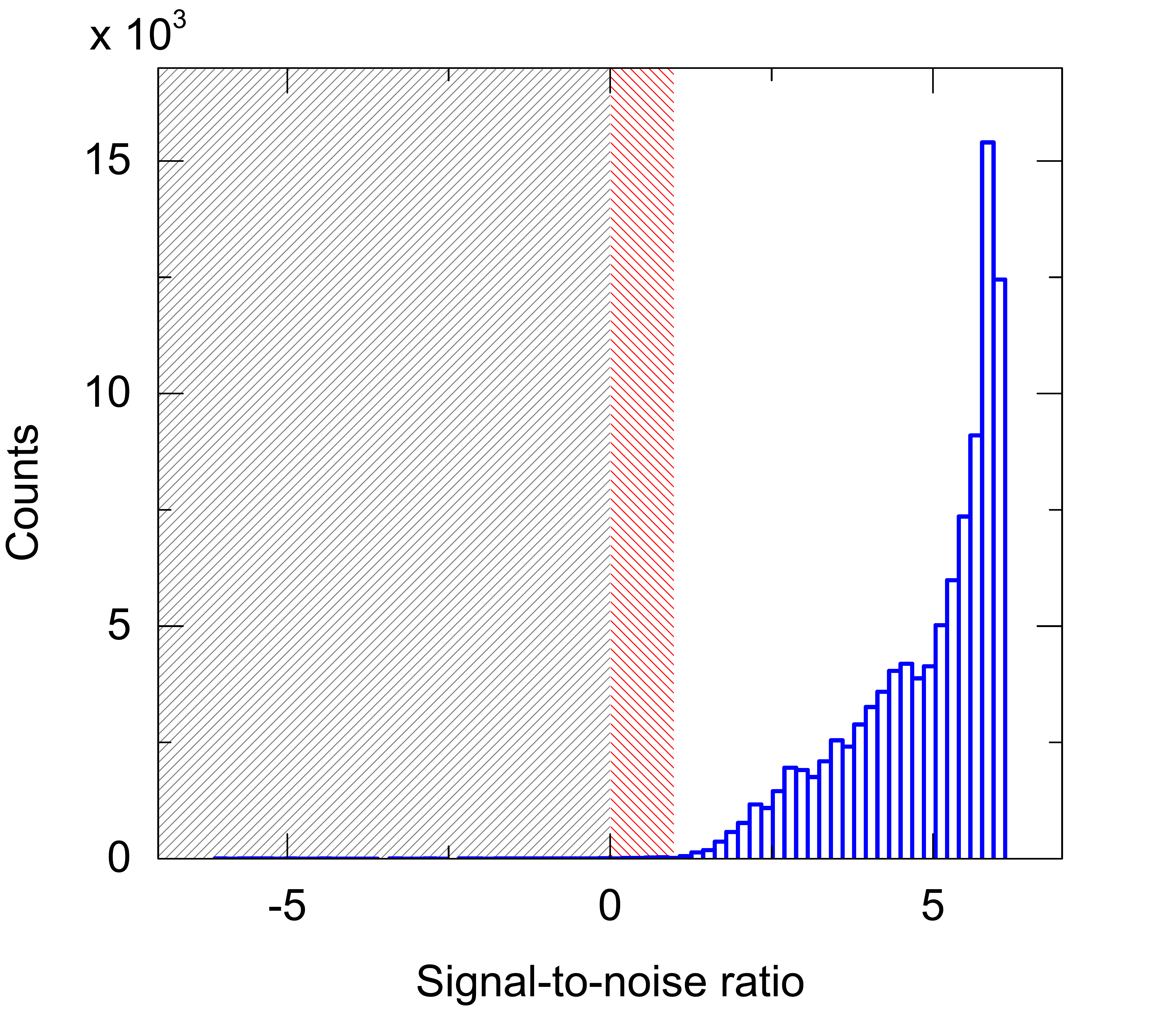}
\caption{(color online) Histogram of the signal-to-noise ratio from $10^5$ trials. The initial conditions are generated randomly from a Boltzmann distribution at 0.4~K. The grey shaded region represents the erroneous detection events and the red shading marks the region where the signal-to-noise ratio is smaller than 1.}
\label{fig:histo}
\end{figure}

To calculate the fidelity of the entire procedure and to verify the analytic results above, we randomly generate $10^5$ initial conditions for a spin-up electron from the Boltzmann distribution. Going through the driving and amplification procedures, we obtain the in-phase amplitude and compute the signal-to-noise ratio. Finally as the noise obeys a Gaussian amplitude distribution \cite{PNAS1986b}, the probability of predicting the correct result is calculated for each trial. Averaging the probability over a large number of trials, the fidelity of the scheme can be derived. Fig.~\ref{fig:histo} shows a histogram of the signal-to-noise ratio for $10^5$ trials, with driving and amplification parameters the same as in Fig.~\ref{fig:time_evo}. It shows that for the vast majority of the cases, the signal-to-noise ratio is greater unity. We extract a fidelity of $99.7\%$ with $\gamma t_\mathrm{det}=4$.

\section{conclusion}\label{conclusion}
In conclusion, we propose a scheme for electronic spin readout using an oscillating magnetic field gradient. The scheme is composed of four stages: cooling, driving, amplification and detection. Trap anharmonicities limit the maximum amplitude of the driven motion and hence cooling and amplification stages are required. The cooling process guarantees that the motion of electron is dominated by coherent driving rather than the random thermal motion. Then the motion is amplified so that it can be read out by a circuit at ambient liquid helium temperature. We analyze the electron motion and optimize the procedure, giving readout fidelity estimates of $99.7\%$ within $25~\mu$s ($20~\mu$s driving, $4~\mu$s detection, $<1~\mu$s cooling and amplification), based on well-achievable parameters. The limitation on time and fidelity stems from the magnetic driving process, and further improvements could be made in more favourable experimental conditions. For example, lowering the electron's axial motion temperature $T_0$ to 10~mK using a dilution refrigerator would circumvent the cooling process and use only $8~\mu$s for entire measurement procedure. Supplying a larger $I_\mathrm{drive}$ to the wire for a larger magnetic field gradient would also help. Using higher trap frequencies confining the electron more tightly would further be beneficial to limit effects from anharmonicity.

The trapped electron system outlined in this proposal in theory meets the five DiVincenzo criteria \cite{DiVincenzo} for the physical implementation of a quantum computer which we summarise for convenience below:
\begin{enumerate}
\item The qubit is realized by the electron spin states in a magnetic field, and scalability is possible with surface Paul traps, similar to trapped ion architectures \cite{Wineland1998,Nature2002Wineland}.
\item Spin initialization is accomplished by projective state readout combined with single qubit gates conditioned on the result of the measurement.
\item The coherence time of an electron spin should be comparable to that of Zeeman states of trapped ions, which exceeds 1~s in low-noise experimental configurations \cite{APB2016Ruster}.
\item (i) Single qubit gate can be realized in a transverse magnetic field in about 10~ns \cite{Nature2011Wineland}; (ii) Two-qubit gates can be realized using microwave near-field that couples the spin states to the motion \cite{PRA2000Sorensen,Nature2011Wineland}. The gate operation time depends on the coupling rate, which scales with mass $m$ as $m^{-1/2}$, so the expected gate time for electrons is about 1~$\mu$s, two orders of magnitude shorter than for ions. Both gate times are much shorter than the coherence time.
\item Qubit measurement can be realized as proposed here with a fidelity $>$ 99\% in 25$~\mu$s at 4~K.
\end{enumerate}
Therefore, it seems feasible to build an all-electric quantum computer based on trapped electrons with a gain in speed and robustness as compared to trapped ion approaches.

\begin{appendix}
\section{Field of current-carrying circuits}\label{field}
Here we derive the electric field and magnetic field of the current-carrying circuits shown in Fig. \ref{schematic}(b). We term the two vertical wires near the center of the trap as ``source wires'' and the horizontal wires near the center as ``drain wires''. The wires are of width $w=10~\mu$m, height $1~\mu$m, and the centers of the source wires are separated by $2d=20~\mu$m.
Although the field can be fully simulated, we still start with a few assumptions and derive the analytical solution, simplifying our analysis of imperfection in Appendix \ref{imperfections}.
As the size of the trap is sub-millimeters and the wavelength of a 300 MHz microwave field corresponds to 1 m, it is reasonable to apply a quasi-static approximation. In the analysis of the magnetic field, we assume the wire is a geometric line, while for the electric calculation, we retain its finite thickness, as the field of a charged infinitely-thin wire is ill-defined. The height of wire is roughly twice the skin depth at 300 MHz ($0.4~\mu$m), so the current distribution in the cross section is approximately homogeneous. We also ignore the complex behavior of the current at the source/drain wire junction. With the above assumptions in mind, we now discuss the properties of the fields. Due to the mirror symmetry, the magnetic field of the two drain wires cancels along the axial direction in the trap chip plane, and the magnetic field due to the upper and lower parts of each source wire vanishes at the trapping center $(0,0,h)$. As seen from $(0,y,h)$, the current of each source wire cancels except for a segment of length $2y$. Two segments produce a magnetic field along the $x$-direction
\begin{equation}
B_x(y)=\frac{\mu_0}{4\pi}\frac{2I_\mathrm{drive}hy}{(h^2+d^2)^{3/2}},
\end{equation}
where $I_\mathrm{drive}$ is the current in each drain wire and we assume $y\ll h$. For $I_\mathrm{drive}=1$~A, $h=33~\mu$m and $d=10~\mu$m, the magnetic gradient is $B_x'(y)=160$~T/m. A finite element simulation gives $B_x'(y)=150$~T/m, where the slight discrepancy is due to finite size of wires. Further taking into consideration the shielding effect of the surface electrodes we obtain $B_x'(y)=90$~T/m, assuming the height of the surface electrodes and the thickness of the dielectric material separating electrodes and current-carrying wires are both 1~$\mu$m.

The axial electric is determined by both the absolute voltage on the wires and the voltage difference induced by $I_\mathrm{drive}$. Denoting the absolute voltage at both cross points of source and drain wires $(\pm d,0,0)$ to be $V_0$, a Taylor expansion of the electric potential $\Delta V(y)$ originating from the current-carrying wires yields a quadratic leading term (the constant potential has no effect, the linear term vanishes due to symmetry) $e\Delta V=\eta m\omega^2y^2/2$, with $\eta$ being a dimensionless constant describing the ratio of $\Delta V$ to the axial trap potential. The conductivity of gold at 4~K is $\sigma=4.5\times 10^9$ S/m \cite{CRChandbook} and $V_0=2$~mV (corresponding to the voltage drop of 1~A current through 100 $\mu$m-long wire). We obtain $\eta=-4\times 10^{-3}$. The quadratic potential is oscillating in time, similar to a parametric drive but with frequency $\omega$, so it has little effect on the electron motion, even though the electric force of the wires is much larger than the  magnetic force for the amplitude of the electronic axial motion( $\sim\mu$m).

\section{Imperfections due to asymmetry of current-carrying circuits}\label{imperfections}
\begin{figure}\centering
\renewcommand\thefigure{A1}
\includegraphics[width=0.48\textwidth]{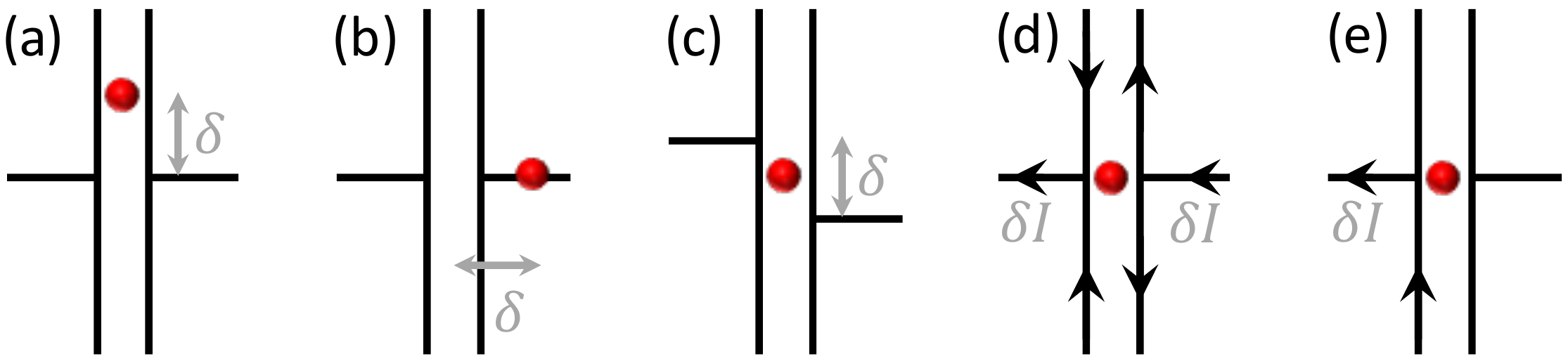}
\caption{(color online) Illustration of experimental imperfections. The correspondence to discussion in Appendix B is (a) $\leftrightarrow$ i, (b) $\leftrightarrow$ ii, (c) $\leftrightarrow$ iii, (d) $\leftrightarrow$ iv and v , (e) $\leftrightarrow$ vi.}
\label{fig:A1}
\end{figure}

In this section we estimate the residual axial electric and magnetic field arising from experimental imperfections that break the symmetry of the circuits. As the driving frequency is 300~MHz and the Zeeman splitting can be made sufficiently different, the effect of the static magnetic field in our readout scheme can be eliminated. Still, our design may also be used for microwave gates, where the magnetic field can cause off-resonant spin flips or an AC Zeeman shift, so we calculate the magnetic field explicitly. The imperfections considered are as follows:

i. $y$-displacement of the trap center (Fig. \ref{fig:A1}(a)).

ii. $x$-displacement of the trap center (Fig. \ref{fig:A1}(b)).

iii. $y$-misalignment of the two circuits (Fig. \ref{fig:A1}(c)) ($x$-misalignment only shifts $d$).

iv. Phase difference $\delta\psi$ between currents in two the circuits. The out-of-phase current is shown in Fig. \ref{fig:A1}(d) with $\delta I=I\delta\psi/2$.

v. Amplitude difference between currents in the two circuits. It is effectively the same as a phase difference so it is also illustrated by Fig. \ref{fig:A1}(d), and $2\delta I$ is the current difference.

vi. Resistance asymmetry within one circuit. It leads to an unbalanced current in upper and lower source wire (Fig. \ref{fig:A1}(e)).

electric and magnetic field originating from the imperfections are presented in table A1.

\begin{table}\centering
\renewcommand\thetable{A1}
\newcommand{\tabincell}[2]{\begin{tabular}{@{}#1@{}}#2\end{tabular}}
\begin{tabular}{|c|c|c|c|}
\hline
\tabincell{c}{Imperfection\\ type }& \tabincell{c}{Imperfection \\ value} & $eE_y/\mu_\mathrm B B'$ & magnetic field ($\mu$T)\\
\hline
i & $\delta=0.1~\mu$m & 1.5 & 8\\
\hline
ii & $\delta=0.1~\mu$m & 0 & 16\\
\hline
iii &  $\delta=0.1~\mu$m & 0 & 4\\
\hline
iv & $\delta\psi=2\pi/1000$ & 0 & 14\\
\hline
v & $\delta I=I/1000$ & 0 & 2\\
\hline
vi & $\delta I=I/1000$ & 0.1 & 3.6\\
\hline
\end{tabular}
\caption{Effects of different types of imperfections on axial electric and residual magnetic field.}
\end{table}

With realistic requirements on precision in fabrication, our design successfully reduces the electric force to the same order as the magnetic gradient force. Spin-echo sequences can be applied to further reduce the force: resonantly driving the spin qubit while modulating $I_\mathrm{drive}$ at the qubit Rabi frequency $\Omega_0$, i.e. $I_\mathrm{drive}(t)=I_\mathrm{drive}(0)\exp{i\Omega_0 t}$, ensures that the spin-dependent magnetic gradient force always excites the axial motion but the spin-independent electric the magnetic field average to zero. The motion driven by the electric should be much smaller than that driven by the magnetic gradient, such that Rabi frequency $\Omega_0\gg eE/(\mu_\mathrm BB' t_\mathrm{drive})$, where $t_\mathrm{drive}=40~\mu$s is the driving time. For our design, only $\Omega_0\approx1$~MHz is required, but for the common straight wire design, with $eE/(\mu_\mathrm BB')\approx10^3$, Rabi frequencies in the GHz range are needed, which is practically challenging.

\section{Signal-to-noise ratio}\label{stn}
In this section we derive the signal-to-noise ratio following Refs. \cite{RMP1986,PNAS1986b}. First we analytically derive the expectation value of noise registered by the phase-sensitive detector.
We consider Johnson noise as the only noise in the system.
We start with the equation of motion
\begin{equation}
\ddot y+\gamma \dot y +\omega^2 y=-\frac{e}{md_\mathrm{eff}}v,
\end{equation}
where $d_\mathrm{eff}=V/E_y\approx 66~\mu$m for our case and $v$ is the voltage applied by the circuit. The corresponding Green's function is defined as
\begin{equation}
\left( \frac{\mathrm d^2}{\mathrm dt^2} +\gamma\frac{\mathrm d}{\mathrm dt}+\omega^2\right)G(t-t')=\delta(t-t'),
\end{equation}
which is solved in Ref. \cite{RMP1986}:
\begin{equation}
G(t-t')=\frac{\theta(t-t')}{\omega}e^{-\gamma(t-t')/2}\sin\omega(t-t'),
\end{equation}
where $\theta$ is the step function.
Assuming the interaction between the electron and detection circuit starts at $t=0$, the random motion of the electron can be expressed using the Green's function
\begin{equation}
y_\mathrm N(t)=-\frac{e}{md_\mathrm{eff}} \int_0^t G(t-t_1)v_\mathrm N(t_1) \mathrm d t_1,
\end{equation}
where $v_N$ describes the Johnson noise, whose expectation value vanishes, $\langle v_\mathrm N\rangle=0$, but the correlation is finite, $\langle v_\mathrm N^2\rangle=2k_\mathrm B T_\mathrm e R$.
The voltage across the detection circuit resistance is the sum of the Johnson noise and the electron-induced voltage, so the noise voltage is $V_\mathrm N=v+\gamma m d_\mathrm{eff}\dot y_\mathrm N/e$. Recalling the EOM, the noise voltage is obtained as
\begin{equation}
V_\mathrm N(t)=-\frac{md_\mathrm{eff}}{e}\left(\frac{\mathrm d^2}{\mathrm d t^2}+\omega^2\right)y_{\mathrm N}(t).
\end{equation}
The noise read out from the phase-sensitive detector is
\begin{equation}
N=\frac{1}{t_\mathrm{det}}\int_0^{t_\mathrm{det}} \cos(\omega t)V_\mathrm N(t) \mathrm dt.
\end{equation}
Ignoring the exponential decay terms, such as $\exp(-\gamma t_\mathrm{det})$ and using $\gamma\ll \omega$, after careful calculation we obtain
$\langle N^2\rangle={k_\mathrm B T_\mathrm e R}/{t_\mathrm{det}}$

The damped electron motion induces signal
\begin{equation}
S=\frac{1}{t_\mathrm{det}}\int_0^{t_\mathrm{det}} e^{-\gamma t/2}V_0\cos^2\omega t \mathrm d t\approx V_0/(\gamma t_\mathrm{det}),
\end{equation}
where we use the assumption $\exp(-\gamma t_\mathrm{det})\ll 1$ again and $V_0=\gamma m d_\mathrm{eff} \omega A_0/e$ is the voltage induced by the coherent electron motion at the beginning of the detection. Therefore $S^2=mR(\omega A_0)^2/(\gamma t_\mathrm{det}^2)$, and the signal-to-noise ratio is given as in eq. (\ref{eq:stneq}).

\end{appendix}

\begin{acknowledgments}
The authors would like to thank D. Leibfried for pointing out the problem with the electric field produced by current-carrying wires.
\end{acknowledgments}

\bibliographystyle{apsrev4-1}
\bibliography{Peng2016_references}

\end{document}